\documentclass[a4paper]{spie}  

 \usepackage{subcaption}
\usepackage{amsmath,amsfonts,amssymb}
\usepackage{graphicx}
\usepackage{chemformula}
\usepackage[colorlinks=true, allcolors=blue]{hyperref}
\usepackage{placeins}
\title{Pollux UV \& FUV polarimeters: first lab results}

\author[a]{Adrien Girardot}
\author[a]{Ninon Demelier}
\author[a]{Coralie Neiner}
\author[a]{Jean-Michel Reess}
\author[b]{Juan Ignacio Larruquert}
\author[b]{Paloma Lopez-Reyes}
\author[b]{Alejandro Garcia Canas}
\author[b]{Nuria Guti\'errez Luna}
\author[b]{Ana Iglesias Moreno}
\affil[a]{LIRA, Observatoire de Paris, Université PSL, Sorbonne Université, Université Paris Cité, CY Cergy Paris Université, CNRS,92190 Meudon, France}
\affil[b]{GOLD-IO-CSIC, Instituto de \'Optica - Consejo Superior de Investigaciones Cient\'ificas, Madrid, Spain}

\authorinfo{{Further author information: (Send correspondence to A.G.)\\A.G.: E-mail: adrien.girardot@obspm.fr}}

\begin{document} 
\maketitle

\begin{abstract}
Pollux is a high-resolution spectropolarimeter proposed by a European consortium for the Habitable Worlds Observatory (HWO). Its design covers a broad spectral range from the far-ultraviolet (FUV) to the near-infrared (97-1\,750 nm), with polarimetric channels relying on \ch{MgF2} birefringent optics in the mid- and near-UV (MUV-NUV), and on an innovative all-reflective polarimeter in the FUV, where no birefringent material is available. To validate these polarimeters, whose required polarimetric precision is $10^{-3}$, a dedicated vacuum ultraviolet test bench has been developed, with two configurations: one for the MUV-NUV range (120-290 nm) and one for the FUV range (98-120 nm). We present the first laboratory results obtained with this bench. On the MUV-NUV configuration, the full optical chain has been integrated: a first polarised spectrum of the deuterium lamp was acquired, the polarisation generation subsystem was validated against Mueller matrix predictions, and a first end-to-end polarimetric measurement was performed. On the FUV configuration, the windowless deuterium plasma source has been characterised, the alignment strategy of the K-mirror modulator has been implemented, and the mirror-based analyser has been manufactured and tested, showing a polarisation extinction ratio of 10 at 120 nm. These results demonstrate the operation of the bench and pave the way for the characterisation of the polarimetric precision of the Pollux polarimeters, increasing the Technology Readiness Level of UV spectropolarimetry for HWO.
\end{abstract}



\section{Introduction}

Pollux is a high-resolution spectropolarimeter proposed by a European consortium as a potential instrument aboard the Habitable Worlds Observatory (HWO). HWO is a flagship mission currently under study, with five telescope concepts being investigated, featuring apertures between 6 and 8 meters and a planned launch after 2040. The current design of Pollux covers a broad spectral range, from 97 to 1\,750 nm, with a spectral resolution up to $R=100\,000$ \cite{neiner_pollux_2026}. In particular, it includes three channels in the ultraviolet (UV), extending into the far-ultraviolet (FUV, 97--123 nm), a spectral region of strong astrophysical interest but technically challenging to access. For the near-UV (NUV) and mid-UV (MUV) channels, the polarimeters rely on waveplates and prisms made of \ch{MgF2}, a birefringent material. Below $\sim$118 nm, however, no birefringent material is available, and the FUV polarimeter is therefore based on an innovative all-reflective design \cite{girardot_design_2024, LeGalUnknownTitle2019}.

To validate these polarimeters and increase the Technology Readiness Level (TRL) of UV spectropolarimetry, a dedicated vacuum ultraviolet test bench has been developed at LIRA (Observatoire de Paris). Its architecture has been presented in previous works \cite{girardot2026pollux}: it features two configurations, one optimised for the MUV-NUV bands (120--290 nm) and one adapted to the FUV range (98--120 nm). The objective of this bench is to characterise the polarimetric precision of the Pollux polarimeters, for which the required precision is $10^{-3}$.

This paper focuses on the first laboratory results obtained with the test bench, both for MUV-NUV and FUV polarimetry. Section \ref{sec:muv} presents the MUV-NUV bench, its integration, and the first spectropolarimetric measurements. Section \ref{sec:fuv} presents the status of the FUV bench, including the K-mirror alignment strategy and the manufacturing and testing of the FUV analyser.

\section{MUV-NUV bench design and first tests [120-290 nm]}\label{sec:muv}

\subsection{Bench overview}

\begin{figure}[htbp]
	\centering
	\includegraphics[width=\textwidth]{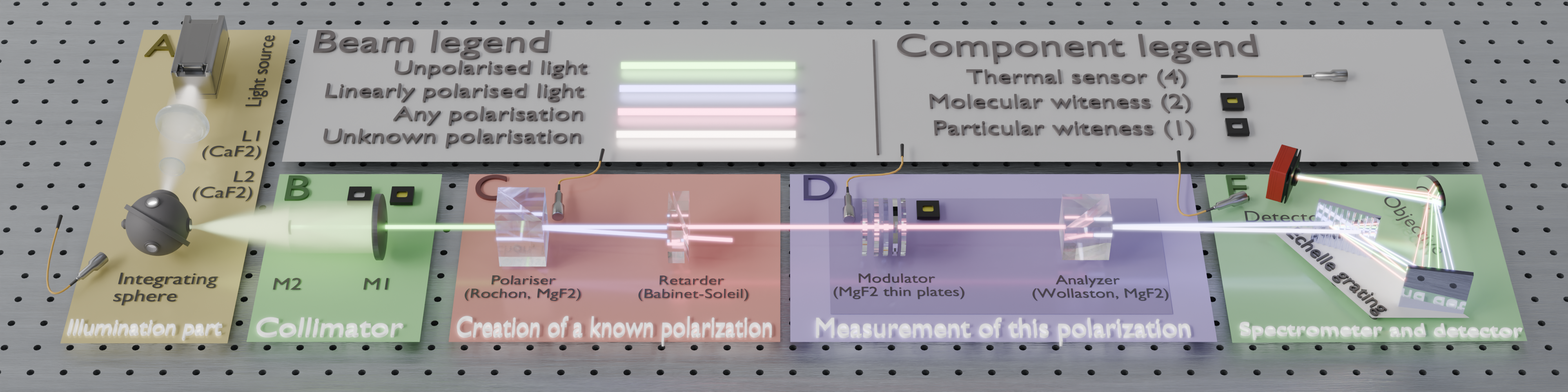}
	\caption{Schematic layout of the MUV-NUV test bench}
	\label{fig:schemaCasstor}
\end{figure}

The MUV-NUV configuration of the test bench is dedicated to the characterisation of the CASSTOR polarimeter, a demonstrator of the MUV-NUV Pollux polarimeters. The experimental setup is composed of five blocks, shown in Fig.~\ref{fig:schemaCasstor}. Block A generates a point source of unpolarised light: the beam of a deuterium lamp, emitting from 115 to 300 nm, is injected via two \ch{MgF2} lenses into an integrating sphere that depolarises the light. Block B is a two-mirror collimator that images the 1 $\mu$m pinhole located at the output of the sphere to infinity, producing a collimated, unpolarised beam of 6 mm diameter. Block C generates any desired polarisation state by combining a \ch{MgF2} Rochon prism with a Babinet-Soleil compensator acting as a continuously tuneable retarder. Block D is the polarimeter under test, composed of a modulator (two pairs of \ch{MgF2} retardation plates) followed by a Rochon prism analyser. Finally, block E is a cross-dispersed echelle spectrometer, operating over diffraction orders 19 to 42 with a resolution of $R \approx 39\,000$, that focuses the dispersed beams onto a Teledyne CIS120 detector.

Since wavelengths below 200 nm are absorbed by air, the whole bench operates inside a vacuum chamber located in a clean room. A two-stage pumping system (primary and turbomolecular pumps) brings the chamber to a secondary vacuum of $10^{-6}$ mbar. The alignment strategy consists in aligning the optical subsystems with the chamber open, and limiting in-vacuum adjustments to fine-tuning steps performed with vacuum-compatible piezoelectric motors. Molecular and particulate contamination, particularly critical in the VUV where deposited molecules can be polymerised by UV radiation, is regularly monitored using contamination witnesses. Thermal measurements show that, once closed and pumped, the chamber provides a very stable environment for the optics (temperature variations below 0.05$^{\circ}$C on the collimator), the main thermal sensitivity being a defocus term that is compensated by a final focus adjustment at operational temperature.

\subsection{First light and polarised spectrum of the lamp}

\begin{figure}[htbp]
	\centering
	\includegraphics[width=0.7\textwidth]{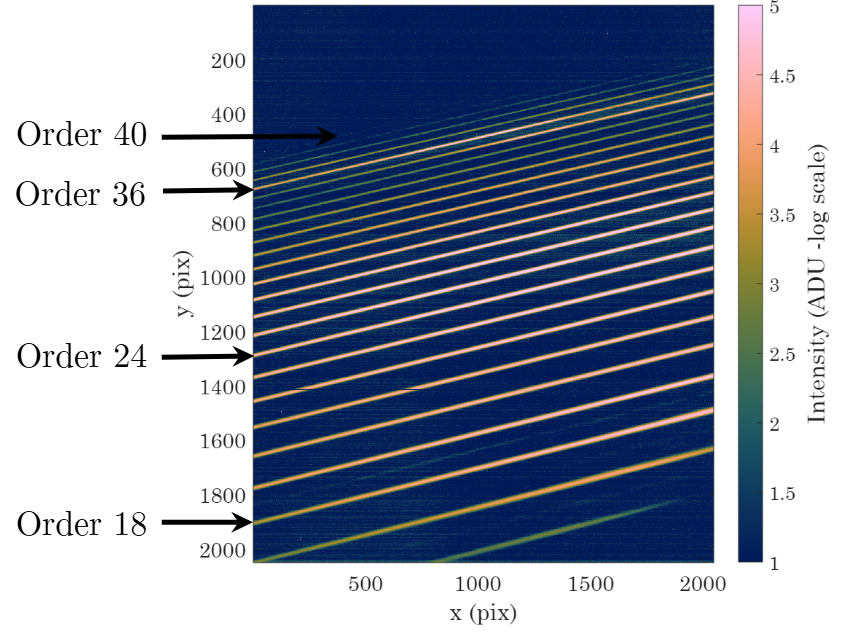}
	\caption{First light of the MUV-NUV test bench. The image is the average of 15 exposures of 1.7 s. Diffraction orders 19 to 41 are visible, covering the spectral range 138--290 nm.}
	\label{fig:firstlightCasstor}
\end{figure}

After the assembly and alignment of blocks A, B, and E, the first light of the bench was acquired with the polarisation blocks removed from the optical path. The resulting image, processed by a dedicated data reduction pipeline (frame averaging, offset subtraction, defective pixel masking, order detection, and spectral extraction), is shown in Fig.~\ref{fig:firstlightCasstor}. Diffraction orders 19 to 41 are visible, covering the spectral range 138--290 nm, in good agreement with the theoretical layout of the spectrometer.

\begin{figure}[htbp]
	\centering
	\includegraphics[width=\textwidth]{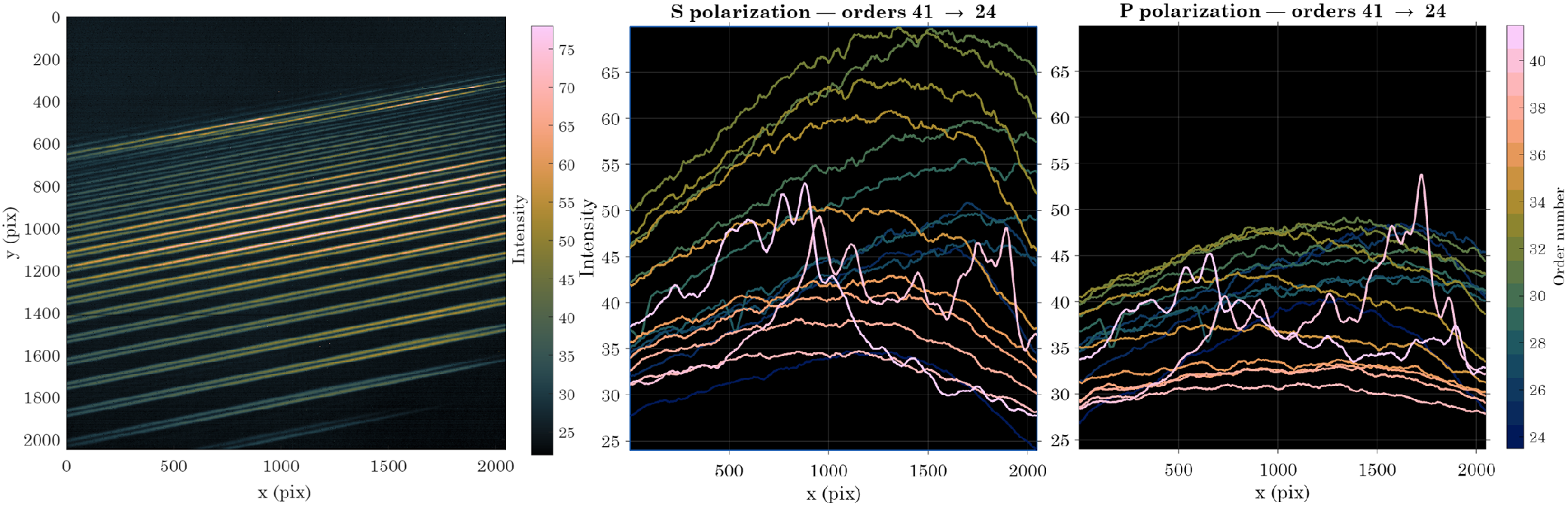}
	\caption{Polarised spectrum of the lamp obtained with the MUV-NUV test bench. Each diffraction order is split by the Rochon prism into two orthogonally polarised components of comparable intensity.}
	\label{fig:polarisedSpec}
\end{figure}

Then, to obtain a polarised spectrum of the lamp, the Rochon prism of the polarimeter was inserted in the beam just before the spectrometer. The prism splits the incoming beam into two orthogonally polarised components, forming two parallel echelle patterns on the detector (Fig.~\ref{fig:polarisedSpec}). This spectrum is used as a reference for the polarimetric measurements of the Pollux polarimeter. A noteworthy result concerns the polarimetric efficiency of the gratings: the two polarised orders exhibit comparable intensities throughout the operating range, indicating that the combined efficiency of the two gratings is approximately balanced for the two orthogonal polarisations. This was not guaranteed by design, as manufacturers rarely provide polarised efficiency data in the VUV, and it is a favourable property for the polarimetric measurements.

\subsection{Polarisation creation (Block C)}

\begin{figure}[htbp]
	\centering
	\includegraphics[width=\textwidth]{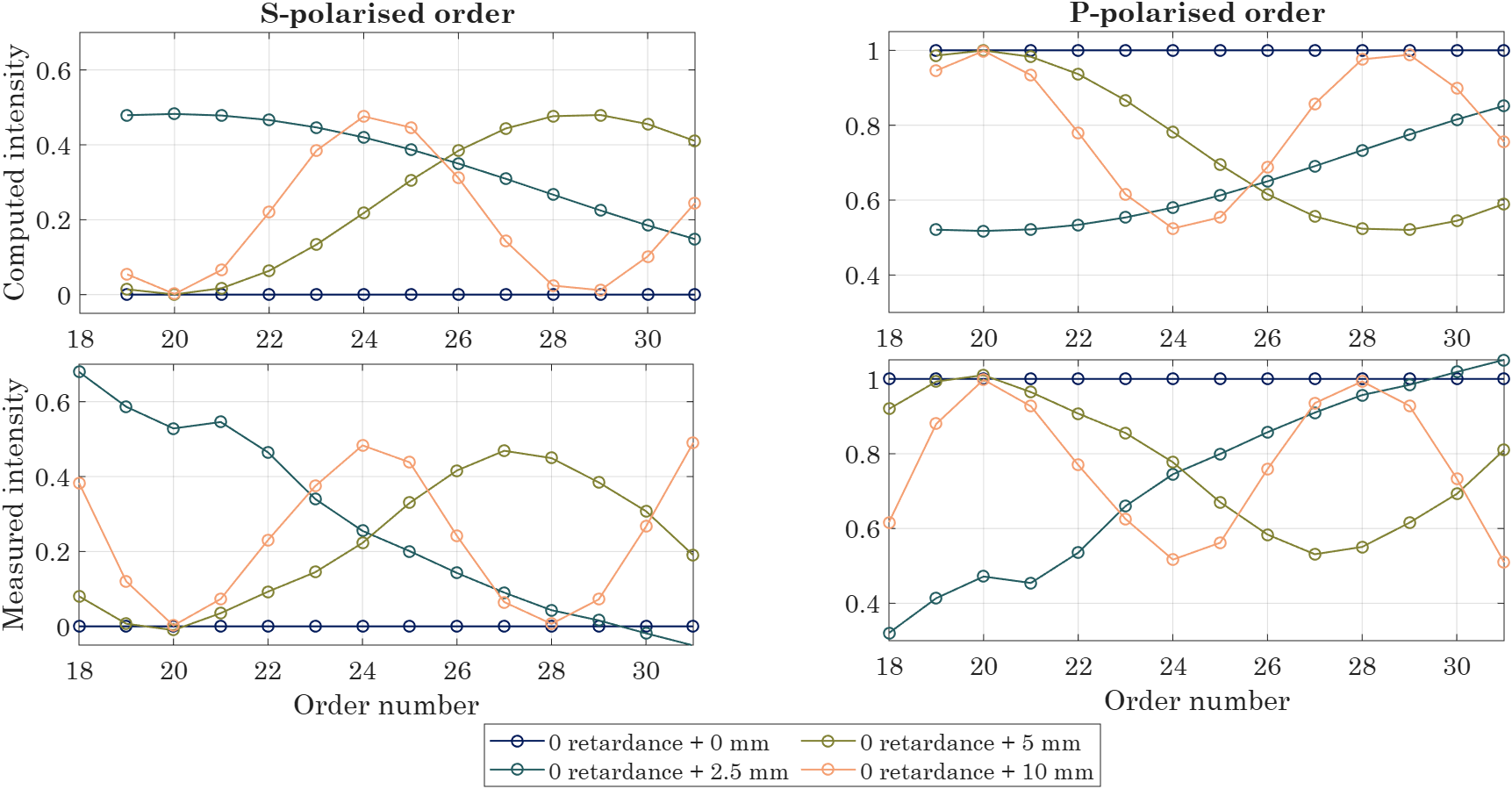}
	\caption{Comparison of theoretical and measured intensities as a function of the Babinet-Soleil position, with the Rochon prism set for P-polarised light. \textbf{Top row:} computed intensities. \textbf{Bottom row:} measured intensities. \textbf{Left:} S-polarised order. \textbf{Right:} P-polarised order.}
	\label{fig:blocC}
\end{figure}

Block C was then integrated to create controlled polarisation states. The Rochon prism transmits a linearly polarised beam, and the Babinet-Soleil compensator, whose zero-retardance position was calibrated beforehand, transforms it into any desired polarisation state as a function of the position of its motorised wedge. The intensity of the different spectral orders was measured for several retardance values and compared to the theoretical values obtained by Mueller matrix calculations (Fig.~\ref{fig:blocC}). The results show a good agreement between the measurements and the theoretical predictions, validating our polarisation creation method.

\subsection{First polarimetric measurement}

\begin{figure}[htbp]
	\centering
	\includegraphics[width=\textwidth]{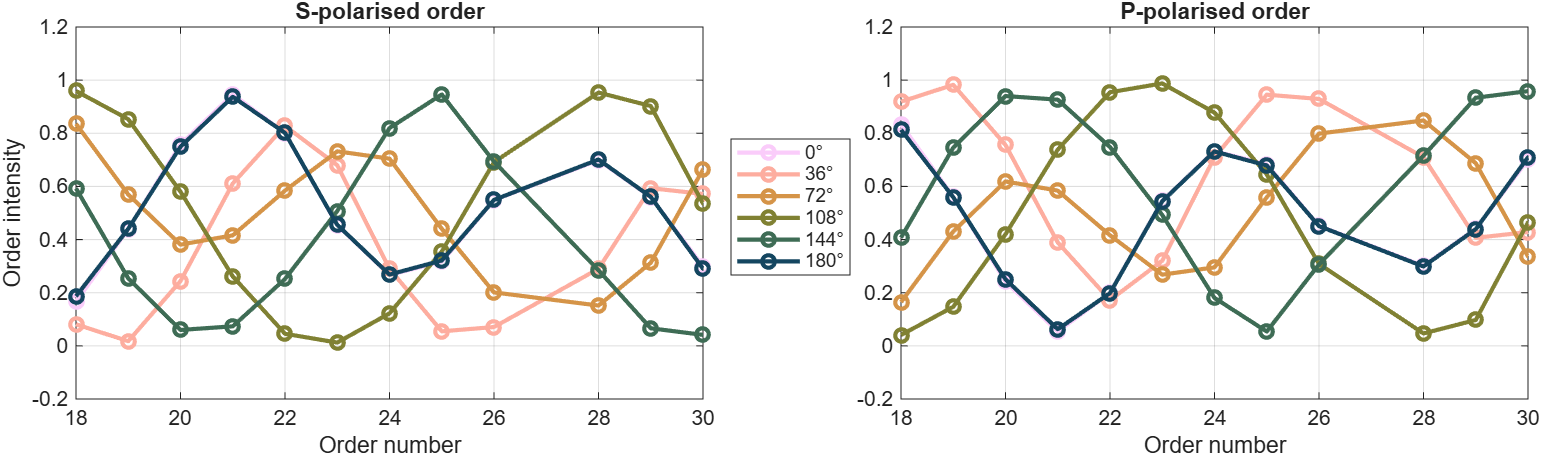}
	\caption{Order intensity modulation as the modulator rotates around the optical axis. Each curve represents a modulator position.}
	\label{fig:order_modu}
\end{figure}

Finally, the MUV-NUV modulator was mounted to perform a first full polarimetric measurement. The measurement consists of six images taken at six modulation angles. For each angle, the intensity of each spectral order is extracted with the data reduction pipeline (Fig.~\ref{fig:order_modu}), and the Stokes vector of the input beam is recovered by demodulation.

\begin{figure}[htbp]
	\centering
	\includegraphics[width=0.7\textwidth]{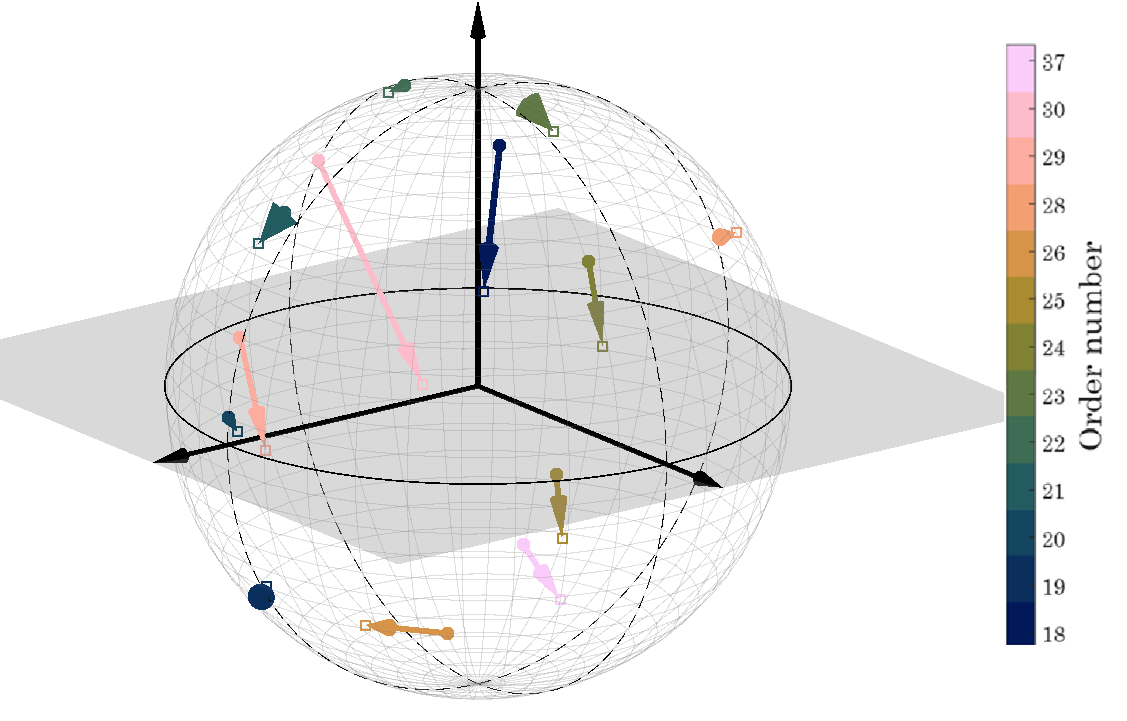}
	\caption{Poincaré sphere representation of the Stokes reconstruction error. For each order, an arrow connects the theoretical Stokes vector to the measured one.}
	\label{fig:sphereresults}
\end{figure}

The measured Stokes vectors are compared, order by order, with the theoretical ones computed with the Mueller formalism. The reconstruction can be visualised on the Poincaré sphere (Fig.~\ref{fig:sphereresults}): for each order, a vector is drawn from the theoretical Stokes vector to the measured one, so that a perfect reconstruction reduces to a single point whereas any error appears as an arrow. This representation provides a useful diagnostic in the case of large errors: for instance, one order shows significant depolarisation, its arrow pointing from the surface of the sphere towards its centre, which may indicate insufficient separation between the S and P beams for that order. These first results validate the complete measurement chain of the bench, from polarisation generation to demodulation, and constitute the starting point for the characterisation of the polarimetric precision of the MUV-NUV Pollux polarimeter.

\section{FUV bench design and first tests [98-120 nm]}\label{sec:fuv}

\subsection{Bench overview}

\begin{figure}[htbp]
	\centering
	\includegraphics[width=\textwidth]{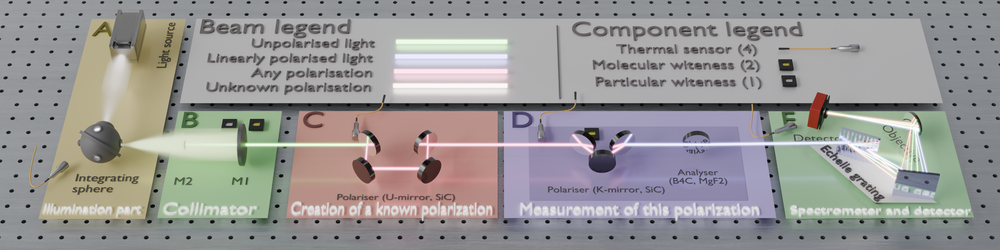}
	\caption{Schematic layout of the FUV test bench}
	\label{fig:schemapollux}
\end{figure}

The FUV configuration of the bench (Fig.~\ref{fig:schemapollux}) is designed to characterise the mirror-based Pollux FUV polarimeter, addressing the absence of birefringent materials in this spectral region. Since no lamp window is transparent in the FUV, the light source is a windowless deuterium plasma generated directly inside the vacuum chamber, allowing emission below 115 nm. The gas flow is injected continuously and evacuated by the pumping system; with a 200 $\mu$m pinhole and a pumping speed of 300 L/s, the chamber pressure remains around $1.6\times10^{-5}$ mbar, compatible with the operation of the bench. The lamp spectrum was measured with the 9 m Rowland-circle spectrometer of the Paris Observatory, and deuterium was selected over nitrogen as it provides more intense emission lines in the FUV range and enables cross-calibration with the MUV-NUV bench. The collimator of the MUV-NUV configuration is reused, as its optical performance remains acceptable down to 100 nm.

The FUV polarimeter under test is composed of a rotating K-mirror acting as modulator and a multilayer mirror acting as analyser. The alignment of the K-mirror is critical to reach high polarimetric precision.

\subsection{FUV light source}\label{sec:fuvlamp}
 
The deuterium lamp used in the MUV-NUV configuration cannot be used in the FUV, as its \ch{MgF2} window is not transparent below $\sim$115 nm. The FUV source must therefore be windowless: the deuterium discharge takes place directly inside the vacuum chamber, without any window separating it from the rest of the system. This requires a continuous flow of deuterium gas injected into the discharge cell and evacuated by the pumping system, while maintaining the pressure difference between the cell and the chamber: the discharge operates at a pressure $P_1 \approx 5\times10^{-1}$ mbar, whereas the chamber must remain below $10^{-5}$ mbar. This differential pumping is achieved by connecting the two volumes only through the source pinhole. A conductance calculation, balancing the flow through the pinhole against the pumping speed of the turbomolecular pumps ($\approx$300 L/s for deuterium), shows that with a 200 $\mu$m pinhole the chamber pressure stabilises around $1.6\times10^{-5}$ mbar, compatible with the operation of the test bench. The source is cooled by a closed-loop water-cooling system driven by a temperature-controlled Arduino board.
 
\begin{figure}[htbp]
	\centering
	\includegraphics[width=0.3\textwidth]{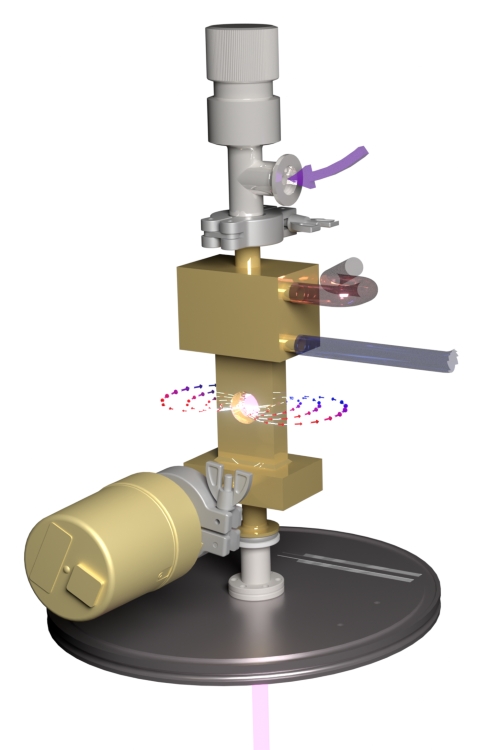}
	\caption{Schematic of the windowless FUV deuterium source}
	\label{fig:SourceFUV}
\end{figure}
 
\begin{figure}[htbp]
	\centering
	\begin{subfigure}[b]{0.45\textwidth}
		\includegraphics[width=\textwidth]{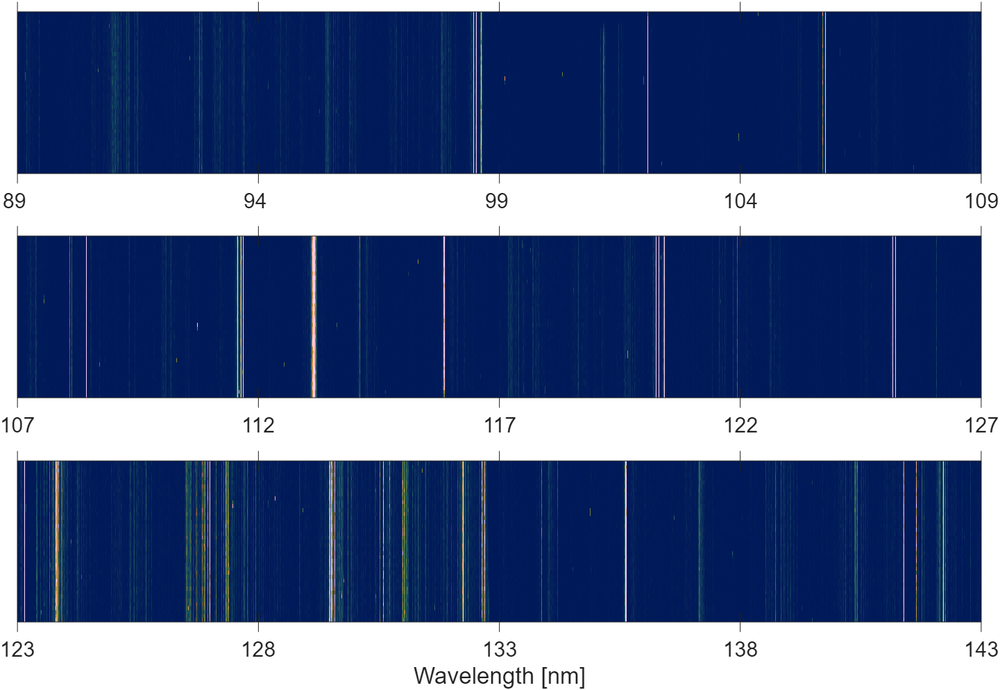}
		\caption{Nitrogen spectrum}
	\end{subfigure}
	\hfill
	\begin{subfigure}[b]{0.45\textwidth}
		\includegraphics[width=\textwidth]{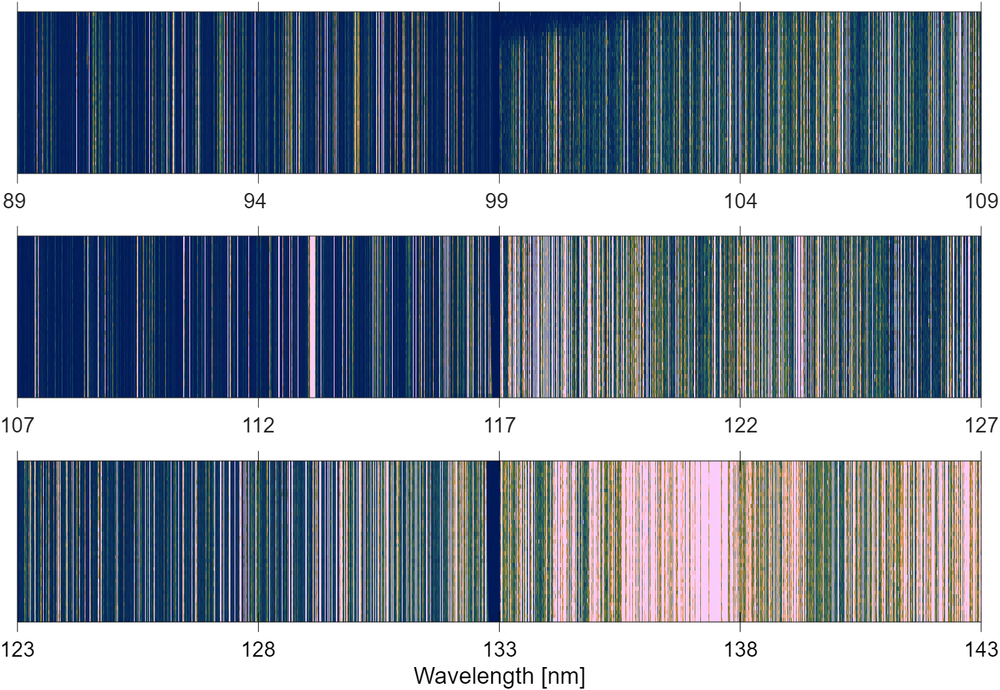}
		\caption{Deuterium spectrum}
	\end{subfigure}
	\caption{Spectra of the FUV lamp measured with the 9 m spectrometer of the Paris Observatory, in identical conditions and with the same colour scale. The spectra are normalised to the maximum intensity of the nitrogen spectrum.}
	\label{fig:spectredeuterium}
\end{figure}

The spectrum of the lamp was measured between 89 and 142 nm using the 9 m Rowland-circle spectrometer of the Paris Observatory, which provides a spectral resolution of up to 200\,000 and records the radiation on a photostimulated-luminescence screen. Two measurements were performed under identical conditions (resolution 50\,000, source pressure $4\times10^{-1}$ mbar, discharge at 750 V and 400 mA, 25 min exposure): one with a nitrogen flow and one with a deuterium flow, in order to select the most suitable gas for the experiment (Fig.~\ref{fig:spectredeuterium}). Both gases exhibit numerous emission lines over the wavelength range of interest, but deuterium provides more lines with higher intensity in the FUV range. Moreover, deuterium is the gas used in the MUV-NUV lamp, and using the same gas for both sources enables cross-calibration between the two configurations of the bench. Deuterium was therefore selected.
 
\subsection{K-mirror alignment}

\begin{figure}[htbp]
	\centering
	\includegraphics[width=0.8\textwidth]{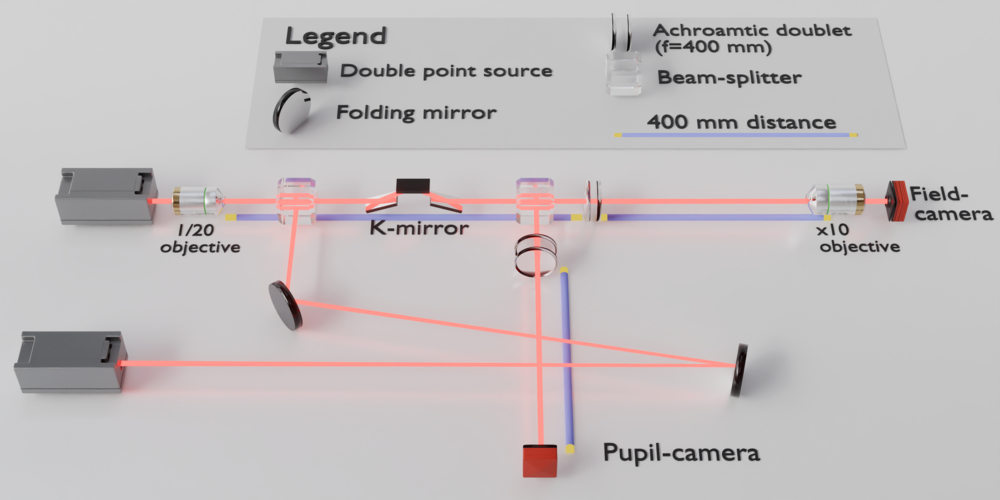}
	\caption{K-mirror alignment method}
	\label{fig:methodealignkmirr}
\end{figure}

The alignment method of the K-mirror is adapted from the method developed for the MICADO K-mirror \cite{thijsak}. A ray entering a perfectly aligned K-mirror along its mechanical axis must emerge along that same axis: aligning the mirror therefore requires merging its optical and mechanical axes. Four degrees of freedom are required: the distance $H$ between the two tilted mirrors and the top mirror, the rotation $\psi$ of the tilted mirrors about the vertical axis, and the tip $\theta$ and tilt $\phi$ of the top mirror. These can be decoupled by imaging two sources simultaneously: a \emph{field} source placed at the entrance of the K-mirror and imaged in a $2f$--$2f$ configuration with two adjacent 400 mm doublets, used to set $H$ and $\psi$; and a \emph{pupil} source placed far from the mirror (about 3 m) and imaged with a 200 mm doublet, used to adjust $\theta$ and $\phi$ (Fig.~\ref{fig:methodealignkmirr}).

When the K-mirror is rotated, the image of a source rotates around a centre which itself rotates around the mechanical axis: the image motion of an on-axis point source is a circle, whereas that of an off-axis point source follows a cardioid. Locating the centre of rotation requires two points, which is why each source is a double source: two lamp-fed fibres clamped 0.9 mm apart on a movable X-Y mount, the brighter point of each pair being the one set on the mechanical axis.

The procedure proceeds in two stages. First, the spot positions are recorded for K-mirror rotations of 0$^{\circ}$, 120$^{\circ}$, and 240$^{\circ}$. From these, a software developed by the MICADO team determines the spot positions, identifies the brightest of each pair, and computes the centres of rotation $C_1$, $C_2$, and $C_3$, which lie on a circle 120$^{\circ}$ apart. The centre $O$ of that circle marks the true mechanical axis, while the diametrically opposite point $C_3'$ is the image of the mechanical axis in the input space for a 240$^{\circ}$ K-mirror angle. Moving the X-Y mounts brings the brightest spots onto $C_3'$, placing them on the mechanical axis. Second, their images are brought onto $O$ using the four degrees of freedom of the K-mirror: the field source is aligned with $H$ and $\psi$, the pupil source with $\theta$ and $\phi$. The image of the mechanical axis in the input space is then merged with the mechanical axis in the output space, and the K-mirror is roughly aligned. Near final alignment, a smaller source separation is used for the field source, obtained by imaging the fibres through a reversed $\times 20$ objective, while a $\times 10$ objective placed in front of the field camera oversamples the Airy disks, allowing a second, finer iteration.

A practical constraint was to design an alignment bench that fits around the vacuum chamber, so that the K-mirror can be aligned without being displaced. A few folding mirrors and two beam splitters allow most of the alignment optics to be installed on a breadboard outside the vacuum chamber, and the chamber can be opened and closed without moving most of the alignment bench.

\begin{figure}[htbp]
    \centering
    \includegraphics[width=0.5\linewidth]{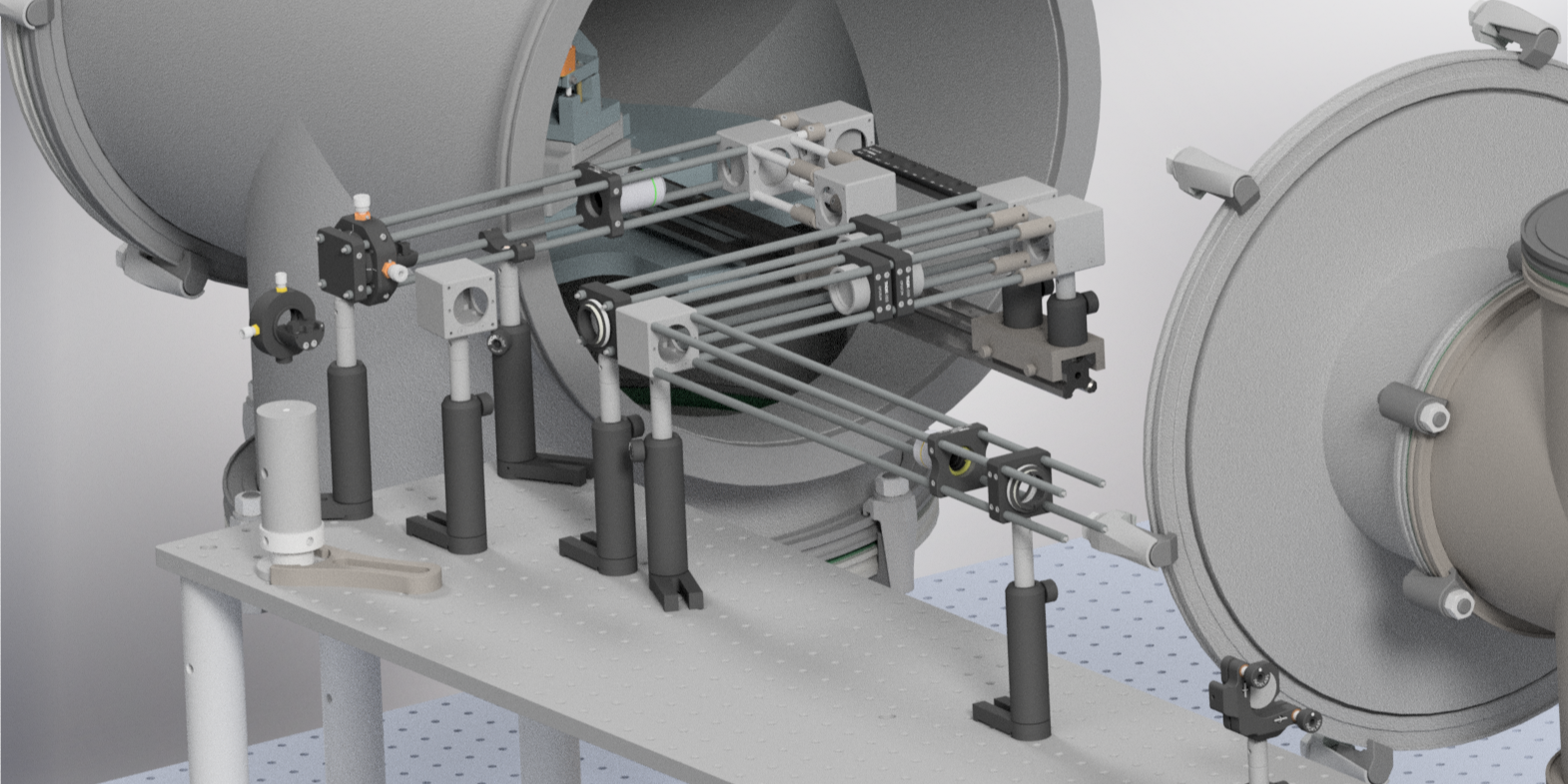}
    \caption{CAO view of the test bench designed to fit around and inside the vacuum chamber}
    \label{fig:placeholder}
\end{figure}

\subsection{Analyser manufacturing and testing}

The FUV analyser consists of a \ch{MgF2} coating deposited on top of a \ch{B4C} layer, itself deposited on a BK7 glass substrate. These two materials require different deposition techniques: \ch{B4C} is deposited by ion beam sputtering, while \ch{MgF2} is deposited by thermal vapour deposition. Both depositions were performed in sequence without breaking the vacuum, in order to avoid contamination between the two layers, using the deposition chamber of the GOLD team at Instituto de Óptica - CSIC (Madrid). A 40 nm \ch{B4C} layer, sufficient to render the layer fully opaque at FUV wavelengths, was first sputtered onto the substrates; the \ch{B4C} target was then replaced by a \ch{MgF2} powder, which was evaporated to deposit a 29 nm \ch{MgF2} layer. Two analysers were manufactured, one with hot-deposited \ch{MgF2} (substrate heated to 250$^{\circ}$C) and one with cold deposition (substrate at room temperature).

\begin{figure}[htbp]
	\centering
	\includegraphics[width=0.64\textwidth]{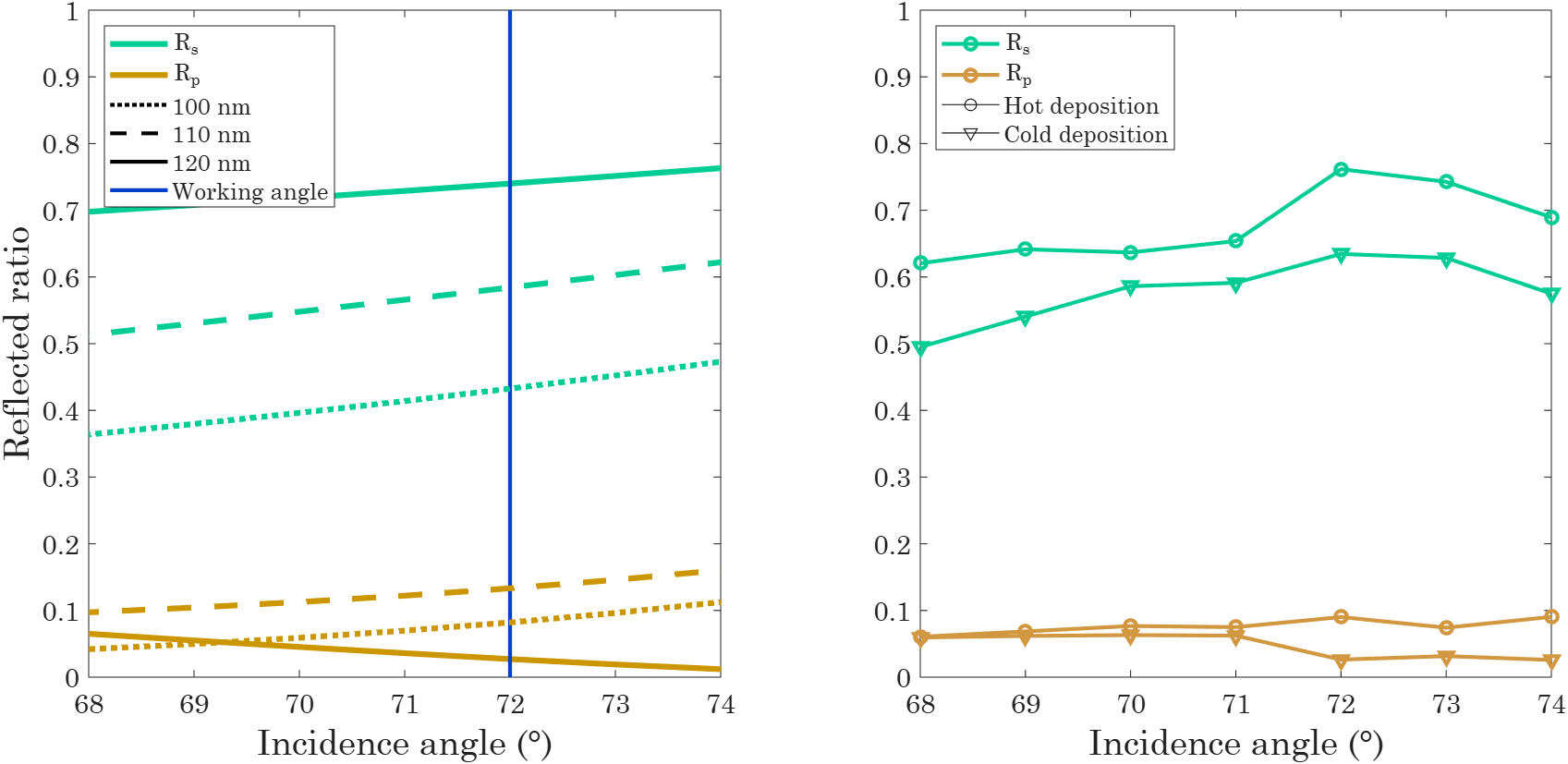}
	\caption{\textbf{Left:} theoretical reflectivity of the analyser for S and P polarisations. \textbf{Right:} measured reflectivity of the analyser at 120 nm.}
	\label{fig:anaresults}
\end{figure}

The polarisation properties of the analyser were then measured at 120 nm on a reflectometer, at the IO-CSIC facility, at different angles of incidence, and compared with theoretical predictions (Fig.~\ref{fig:anaresults}). The results show that the reflectivity for P-polarised light is 10\% of the reflectivity for S-polarised light: the analyser thus has a polarisation extinction ratio of 10 at 120 nm. To confirm these preliminary results, a measurement campaign at a synchrotron facility is planned, which will provide an intense, well-characterised beam with a known and highly stable degree of polarisation, allowing the extinction ratio to be confirmed across the full FUV range.

\FloatBarrier
\section{Conclusion and next steps}

We have presented the first laboratory results of the Pollux test bench, in both its MUV-NUV and FUV configurations. On the MUV-NUV bench, the full optical chain has been integrated: a first polarised spectrum of the lamp was obtained, the polarisation creation subsystem was validated against Mueller matrix predictions, and a first end-to-end polarimetric measurement was performed, with the reconstructed Stokes vectors compared to the theoretical inputs. On the FUV bench, the windowless deuterium source has been characterised, the K-mirror alignment strategy adapted from the MICADO method has been implemented, and the mirror-based analyser has been manufactured and tested, showing a polarisation extinction ratio of 10 at 120 nm.

The next steps will be, on the MUV-NUV side, to consolidate the polarimetric measurements of created polarisation states in order to characterise the polarimetric precision of the MUV-NUV Pollux polarimeter and validate its design; and, on the FUV side, to integrate the K-mirror and the analyser in the vacuum chamber and perform the first FUV polarimetric measurements. This infrastructure represents a critical advancement towards elevating the Technology Readiness Level of UV spectropolarimetric instrumentation, and lays the groundwork for a future implementation aboard the Habitable Worlds Observatory.

\acknowledgements{The authors thank CNES and CSIIT for their financial support.}


\bibliography{BetaManuscrit} 
\bibliographystyle{spiebib} 

\end{document}